\documentclass[%
 aip,
 amsmath,amssymb,
preprint,%
floatfix
]{revtex4-1}

\usepackage{graphicx}
\usepackage{dcolumn}
\usepackage{chemfig}
\usepackage{chemformula}
\usepackage{bm}

\usepackage[utf8]{inputenc}
\usepackage[T1]{fontenc}
\usepackage{mathptmx}
\usepackage{xcolor,colortbl}
\usepackage{xcolor}

\definecolor{Gray}{gray}{0.85}
\definecolor{LightCyan}{rgb}{0.88,1,1}

\newcolumntype{a}{>{\columncolor{Gray}}c}
\newcolumntype{b}{>{\columncolor{white}}c}
\newcommand{\minus}{\scalebox{0.75}[1.0]{$-$}}

\begin{document}

\preprint{}

\title[]{Magneto-ionic modulation of the interlayer exchange interaction in synthetic antiferromagnets}

\author{Maria-Andromachi Syskaki}
\email{msyskaki@uni-mainz.de}
 \affiliation{Singulus Technologies AG, 63796 Kahl am Main, Germany}
 \affiliation{Institut für Physik, Johannes Gutenberg-Universität Mainz, Staudingerweg 7, 55128 Mainz, Germany}
\author{Takaaki Dohi}
\email{takaaki.dohi.e5@tohoku.ac.jp}%
 \affiliation{Institut für Physik, Johannes Gutenberg-Universität Mainz, Staudingerweg 7, 55128 Mainz, Germany}
\affiliation{ 
Laboratory for Nanoelectronics and Spintronics, Research Institute of Electrical Communication, Tohoku University, Japan 
}%
\author{Sergei Olegovich Filnov}
\affiliation{Institut für Physik, Johannes Gutenberg-Universität Mainz, Staudingerweg 7, 55128 Mainz, Germany}
\author{Sergey Alexeyevich Kasatikov}
\affiliation{Helmholtz-Zentrum Berlin für Materialien und Energie GmbH, Hahn-Meitner Platz 1, 14109 Berlin, Germany}
\author{Beatrice Bednarz}
\affiliation{Institut für Physik, Johannes Gutenberg-Universität Mainz, Staudingerweg 7, 55128 Mainz, Germany}
\author{Alevtina Smekhova}
\affiliation{Helmholtz-Zentrum Berlin für Materialien und Energie GmbH, Hahn-Meitner Platz 1, 14109 Berlin, Germany}
\author{Florian Kronast}
\affiliation{Helmholtz-Zentrum Berlin für Materialien und Energie GmbH, Hahn-Meitner Platz 1, 14109 Berlin, Germany}
\author{Mona Bhukta}
\affiliation{Institut für Physik, Johannes Gutenberg-Universität Mainz, Staudingerweg 7, 55128 Mainz, Germany}
\author{Rohit Pachat}
\affiliation{Centre de Nanosciences et de Nanotechnologies, CNRS, Université Paris-Saclay, 10 Boulevard Thomas Gobert, 91120 Palaiseau, France}
\author{Johannes Wilhelmus van der Jagt}
\affiliation{Spin-Ion Technologies, C2N, 10 Boulevard Thomas Gobert, 91120 Palaiseau,France}
\affiliation{Université Paris-Saclay, 3 rue Joliot-Curie, 91190 Gif-sur-Yvette, France}
\author{Shimpei Ono}
\affiliation{Central Research Institute of Electric Power Industry, Yokosuka, Kanagawa 240-0196, Japan}
\author{Dafin\'e Ravelosona Ramasitera}
\affiliation{Centre de Nanosciences et de Nanotechnologies, CNRS, Université Paris-Saclay, 10 Boulevard Thomas Gobert, 91120 Palaiseau, France}
\affiliation{Spin-Ion Technologies, C2N, 10 Boulevard Thomas Gobert, 91120 Palaiseau,France}
\author{J\"urgen Langer}
\affiliation{Singulus Technologies AG, 63796 Kahl am Main, Germany}
\author{Mathias Kläui}
\affiliation{Institut für Physik, Johannes Gutenberg-Universität Mainz, Staudingerweg 7, 55128 Mainz, Germany}
\author{Liza Herrera Diez}
\affiliation{Centre de Nanosciences et de Nanotechnologies, CNRS, Université Paris-Saclay, 10 Boulevard Thomas Gobert, 91120 Palaiseau, France}
\author{Gerhard Jakob}
\email{jakob@uni-mainz.de}%
\affiliation{Institut für Physik, Johannes Gutenberg-Universität Mainz, Staudingerweg 7, 55128 Mainz, Germany}

\date{\today}

\begin{abstract}

The electric-field control of magnetism is a highly promising and potentially effective approach for achieving energy-efficient applications. In recent times, there has  been significant interest in the magneto-ionic effect in synthetic antiferromagnets, primarily due to its strong potential in the realization of high-density storage devices with ultra-low power consumption. However, the underlying mechanism responsible for the
magneto-ionic effect on the interlayer exchange coupling (IEC) remains elusive. In this study, we have successfully identified that the magneto-ionic control of the properties of the top ferromagnetic layer of the synthetic antiferromagnet (SyAFM), which is in contact with the high ion mobility oxide, plays a pivotal role in driving the observed gate-induced changes to the IEC. Our findings provide crucial insights into the intricate interplay between stack structure and magnetoionic-field effect on magnetic properties in synthetic antiferromagnetic thin film systems.

\end{abstract}

\maketitle



\section{Introduction}

Electric field control of magnetism in spintronic devices is considered to be one of the most promising, industrially realizable and energy efficient routes to future applications, for instance in storage, due to its low power consumption \cite{Matsukura2015, Rec_Pro_MRAM, https://doi.org/10.1002/adma.201806662}. Since the first demonstration using a ferromagnetic (FM) semiconductor\cite{Ohno2000}, it has been revealed that various magnetic properties can be modified by an electric field\cite{Chiba2000, Chiba2008}. In metals, electric fields are efficiently screened due to the high charge carrier density, limiting their applications. However, recent spintronic devices are based on ultrathin metal layers with a thickness comparable to the screening length. Accordingly, even in FM metals, where the interface properties dominate the bulk properties, thanks to their ultrathin thickness\cite{GWeisheit2007}, electric field effects have been demonstrated for a variety of magnetic properties such as magnetic anisotropy\cite{Maruyama2009, Endo2010}, Curie temperature\cite{Chiba2011}, damping constant\cite{Okada2014}, proximity effects\cite{Hibino2015,Obinata2015}, exchange interaction\cite{Dohi2016,Ando2016}, interfacial Dzyaloshinskii-Moriya interaction\cite{Kohei2016,Srivastava2018}, and g-factor\cite{Mizuno2022}.  Compared to magnetic semiconductors, magnetic metals are more suitable for industrial applications due to their high Curie temperature, which can be much higher than room temperature. Using applied voltages to control magnetic properties has the potential to pave the way for a new class of computing devices with significantly lower energy consumption\cite{SONG201733}. To be feasible for technological purposes, the changes induced by a voltage must have a significant effect on the magnetic properties of the system. The prospect of producing large effects with minimal voltages has led to a growing interest in magneto-ionic research\cite{Nichterwitz2021AdvancesIM,Scalable,Rojas}.
Ionic liquid gating (ILG)  offers several advantages over solid state gating methods, such as reduced simple device fabrication and large gating areas\cite{Leighton}. Furthermore, the ILs can be engineered to have specific properties, such as low volatility and high thermal stability, enabling the use of this technique in a broad range of operations and environments. The ILG technique also has the potential to be used in conjunction with different classes of advanced materials and device structures, such as two-dimensional materials and nanostructures, to create novel and improved electronic devices\cite{Bauer,Nianpeng}.

Synthetic antiferromagnets (SyAFMs) have gained significant attention due to their unique properties and potential applications in the field of spintronics. The SyAFMs are special systems in that their particular properties are governed by the interlayer exchange coupling (IEC) \cite{RKKY1,RKKY2,RKKY3}, which arises from exchange interactions between magnetic layers through a non-magnetic spacer layer. This interaction is mediated by the conduction electrons in the metal and is dependent on the density of states and distance between the magnetic layers, resulting in either FM or antiferromagnetic (AFM) coupling.
Recent studies have shown the ability to dynamically control the IEC and the switching between FM and AFM coupling of out-of-plane magnetized Co/Pt-based stacks with a Ru interlayer, through magneto-ionic gating with hydrogen ions\cite{Kossak}, ILG\cite{Yang2018,Yang2018_2}, and solid-state Li ion battery technology\cite{Ameziane}. Furthermore, ILG on nanowires, composed of out-of-plane magnetized SyAFM Co/Ni layers with Ru spacer, can reversibly tune the velocity of domain walls and induce large changes in the exchange coupling torque and current-induced domain wall velocity due to the oxidation of the top magnetic layer\cite{Guan2021}.


In our work, we demonstrate significant modulation capability of the SyAFMs IEC by magneto-ionic phenomena, which is linked to the effects of gating on the top FM layer of the SyAFM. We also show that the IEC modulation is also dependent on the thickness of the top layer. This shows the great importance of stack engineering in achieving efficient magneto-ionic control of IEC in SAFs.

\section{Experimental methods}

\begin{figure}[h!]
\centering\includegraphics[width=10cm]{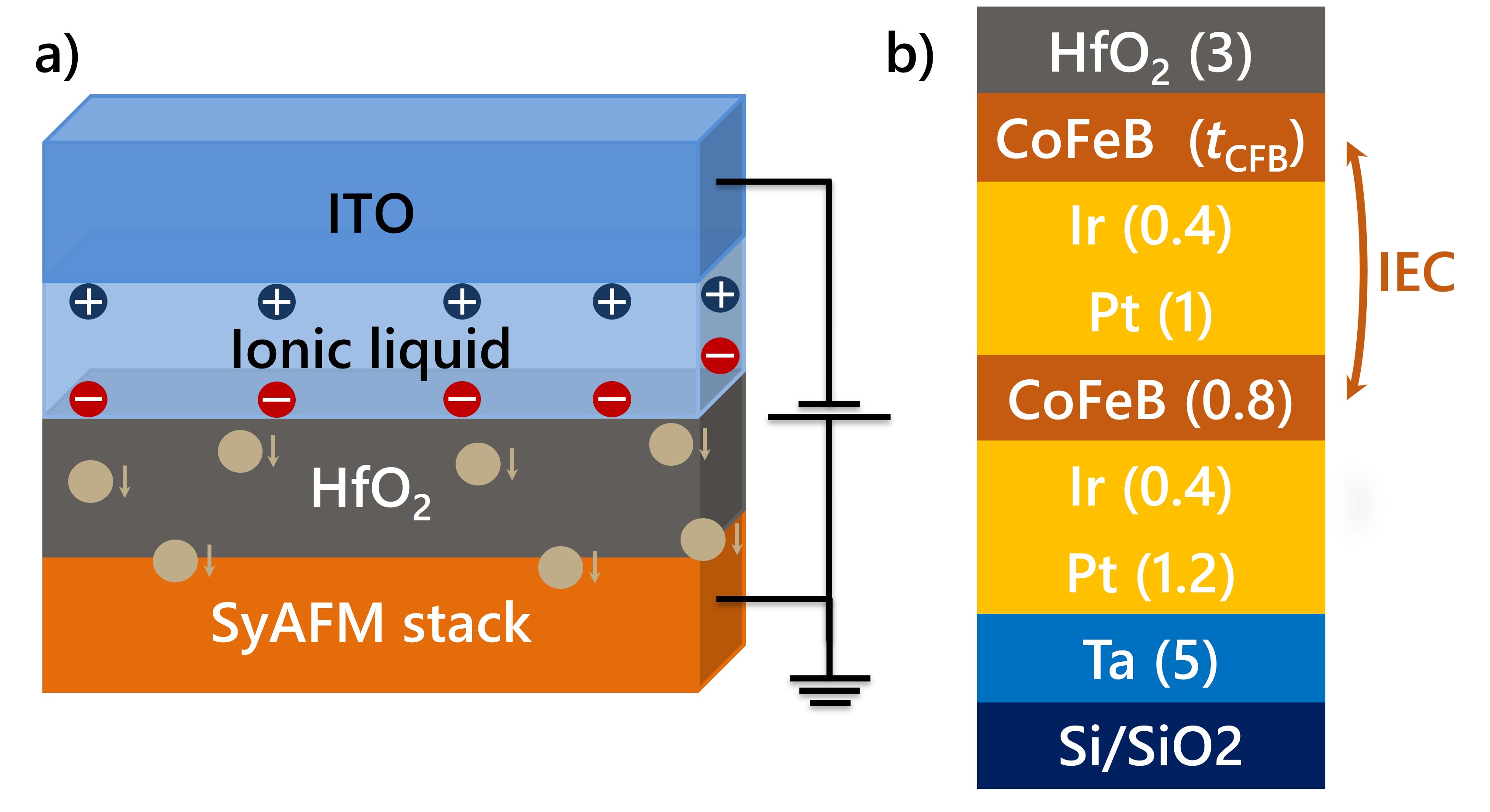}
\caption{\label{fig_Fig1} 
a) A schematic illustration shows the application of a negative gate voltage ($V_\text{G}$<0) with mobile oxygen species (depicted in tan color) moving towards the metallic stack. Figure b) presents the SyAFM stack, where $t_\text{CFB}$ corresponds to the thicknesses of 0.8, 0.9, and 1.0 nm. The numbers in brackets indicate the respective thicknesses in nm.}
\end{figure}

The thin film material stacks were deposited  on thermally oxidized Si/SiO$_2$ substrates at room temperature using a Singulus Rotaris magnetron sputtering tool. This tool offers exceptional reproducibility and sub-Angstrom thickness accuracy, maintaining a base pressure of $5 \times 10^{-8}$ mbar. The metallic layers, including Ta, Pt, Ir, and Co${60}$Fe${20}$B${20}$ (CFB), were grown using DC-magnetron sputtering, while the HfO${2}$ layers were deposited using RF-sputtering from a composite target. The oxide layers function as ionic reservoirs, facilitating magneto-ionic interactions within the metallic stack \cite{PhysRevLett.113.267202, Pt_Co_oxides}.CoFeB was selected as a suitable material due to its compatibility with magnetic random access memory technology in spintronics, owing to its high tunnel magnetoresistance, large perpendicular magnetic anisotropy (PMA), and low damping \cite{Ikeda2010}.
The non-magnetic Ir spacer, with a thickness corresponding to the vicinity of the first peak of the IEC oscillation, mediates the AFM coupling between the FM layers, as depicted in Fig.~\ref{fig_Fig1}b.
Regarding the layer thicknesses, a non-magnetic Ir spacer mediates the interlayer AFM coupling, with a thickness corresponding to the vicinity of the first peak of the IEC oscillation, as shown in Fig.~\ref{fig_Fig1}b, where $t_\text{CFB}$ corresponds to the thickness of 0.8, 0.9 and 1.0 nm of CFB. Here, $t\text{CFB}$ represents the thickness of the CFB layer, which can be 0.8 nm, 0.9 nm, or 1.0 nm.

The as-grown samples exhibit magnetization with a perpendicular easy axis and were not subjected to post-deposition annealing treatment to prevent extensive diffusion of the heavy metals into the adjacent FM layers.
For the implementation of the ILG technique, a sample area of 0.25 cm$^2$ was utilized. The 1-Ethyl-3-methylimidazolium-bis(trifluormethylsulfonyl)imide [EMIM]$^+$[TFSI]$^-$ was used as IL on the surface of the sample\cite{10.1063/1.2898203} and a glass substrate coated with Indium Tin Oxide (ITO) served as the top electrode. An illustrative schematic is presented in Fig.~\ref{fig_Fig1}a. The measurements were conducted at room temperature, and the polar magneto-optical Kerr effect (pMOKE) hysteresis loops were recorded at zero gate voltage ($V_\text{G}$) to exploit the non-volatile nature of the magneto-ionic effects. To investigate the ionic modulation of the IEC field across different material stacks, we set the gating time to a fixed gating time of 30 seconds, enabling a direct comparison of the effect.

\section{Results and discussion}

\subsection{FM thickness dependence of the interlayer exchange coupling field modulation}

Previous studies have clearly demonstrated the AFM-FM transition using ILG, which corresponds to a sign change in the interlayer exchange interaction ($J_\text{int}$) \cite{Yang2018, Guan2021, Kossak}. The interlayer exchange fields, referred to $H_\text{int}$, have commonly been employed as the indicator of the $J_\text{int}$ modulation of \cite{Yang2018_2, Kossak, Ameziane}. In SyAFM stacks, the $H_\text{int}$ field represents the magnetic field at which the AFM-FM transition with a spin flip-like behavior occurs, indicating antiferromagnetic coupling between the two FM layers through the non-magnetic interlayer \cite{Duine2018} at low magnetic fields. However, it should be noted that changes in the total magnetization of the stack also contribute to $J_\text{int}$, which can potentially affect the accurate assessment of the $J_\text{int}$ modulation under gating.

\begin{figure}[h!]
\centering\includegraphics[width=16cm]{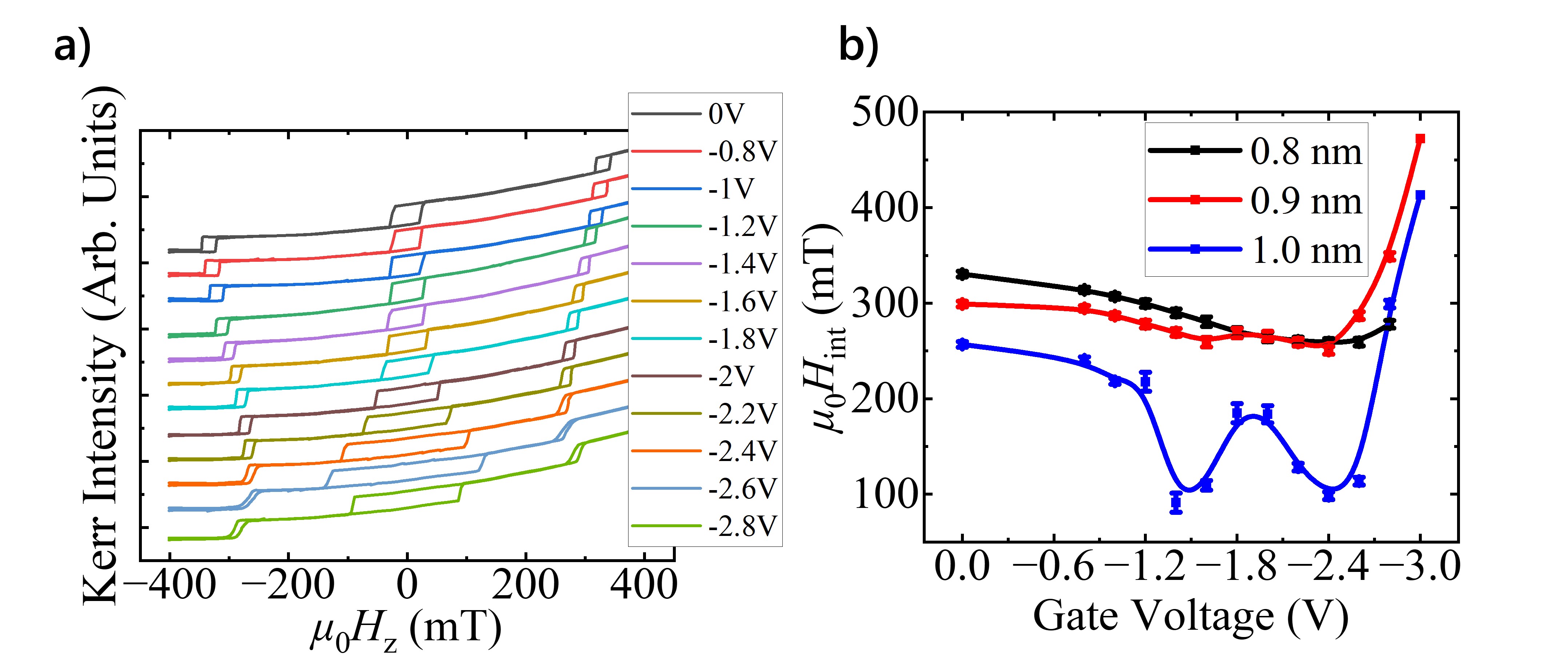}
\caption{\label{fig_syafm} The non-monotonic behavior of IEC field was investigated as a function of magnetic field for a) a top FM layer with a thickness of 0.8 nm of CFB, using pMOKE magnetometry. The non-monotonic behavior of the IEC field was also studied as a function of $V_\text{G}$ for b) different thicknesses of CFB using pMOKE magnetometry. Hysteresis loops were recorded after each 30-second application of V$_\text{G}$.}
\end{figure}

In order to elucidate the influence of the magneto-ionic effect on the modulation of $H_\text{int}$, we conduct a systematic investigation by exploring the relationship between the thickness of the top FM layer and the varying negative $V_\text{G}$. Subsequently, we record the relative changes in the pMOKE measurements following each gating step.
In this investigation, we differentiate between two potential mechanisms: 1) the direct modulation of the Rudermann-Kittel-Kasuya-Yosida (RKKY) interaction\cite{BRUNO1993248}, and 2) the modulation of quantum interference, which is particularly sensitive to the thickness of the FM layer\cite{PhysRevB.52.411}, while the RKKY interaction remains unaffected. In Fig.~\ref{fig_syafm}a, for 0.8 nm CFB, we observe a non-monotonic behavior of the $H_\text{int}$ as a function of the applied magnetic field for progressively negative $V_\text{G}$, corresponding to the migration of mobile oxygen atoms towards the metallic stack.

Our systematic study of the thickness of the top FM layer revealed intriguingly a non-monotonic behaviour of $H_\text{int}$ as a function of $V_\text{G}$, as depicted in Fig.~\ref{fig_syafm}b. When increasing the thickness of the top FM layer from 0.8 nm to 0.9 nm and 1.0 nm, $H_\text{int}$, we observed one peak, two peaks with small amplitude, and two significant peaks, respectively, in the AFM coupling strength regime of the modulation of interlayer exchange interaction $J_\text{int}$. This systematic trend highlights the significant role of the thickness of the top FM in alternating the qualitative behaviour of the magneto-ionic effect on $J_\text{int}$. The observation of this non-monotonic behaviour is supported by superconducting quantum interference device (SQUID) magnetometry measurements presented in Supplementary Material 1 (SM1). The unique nature of IEC is further supported by theoretical calculations\cite{PhysRevB.100.014403, Castro_2020} that indicate the qualitative oscillation of $J_\text{int}$, depending not only on the spacer thickness but also on the thickness of the FM\cite{Oscillation_RKKY}. 

Moreover, the observed dependence of the gate effects on the thickness of the top FM layer, as shown in Fig.~\ref{fig_syafm}b through the behavior of the $H_\text{int}$, indicates that the modulation induced by the magneto-ionic gating does not solely affect the RKKY interaction. Instead, there exists a complex interplay between the RKKY interaction and the modulation of quantum interference\cite{PhysRevB.52.411}. This conclusion is supported by the deviation of $H_\text{int}$ from the conventional trend expected for the RKKY interaction\cite{BRUNO1993248}, as it depends on the distance between magnetic moments rather than solely on the thickness of the FM layer. Therefore, the thickness dependence of $H_\text{int}$ emphasizes the significance of considering the quantum interference modulation when designing synthetic antiferromagnetic (SyAFM) systems.

\subsection{FM thickness dependence of single FM layers}

To gain a deeper understanding of the non-monotonic behavior of $H_\text{int}$, we investigated the influence of the magneto-ionic effect on the modulation of the PMA in the single top FM layer of the previous SyAFM stacks. PMA in FM thin films can be induced by two distinct types of interfaces: one with a heavy metal (HM) and another with an oxide. At the interface with the HM, PMA is induced through spin-orbit coupling\cite{PhysRevB.39.865}, and is significantly influenced by the abruptness, or the thickness transition,at the interface\cite{PhysRevB.62.5794,10.1063/1.3701585}. On the other hand, the origin of PMA at the interface between a FM and an oxide arises from the hybridization between the FM's 3d$_z$ orbitals and the oxide's 2p$_{xy}$ and 2p$_{yz}$ orbitals\cite{Rodmacq,Yang}. Regarding the magneto-ionic effects on the single FM layer, previous studies of Co/HfO$_2$ \cite{Liza} and CoFeB/HfO$_2$\cite{Rohit} systems have demonstrated that under negative V$_\text{G}$, these stacks undergo a transition from in-plane anisotropy to PMA- This transition can be attributed to the modulation of the oxidation level of the FM layer. 

\begin{table}[h!]
    \centering
    \begin{tabular}{|| c || c ||} 
 \hline
    Stack coding & Material stacks \\ [0.5ex]
 \hline\hline
FM Stack \verb|#| 1 & sub// Ta (5.0)/ Pt (1.0)/ Ir (0.4)/ CFB (0.8)/ HfO$_{2}$ (3.0) \\
FM Stack \verb|#| 2 & sub// Ta (5.0)/ Pt (1.0)/ Ir (0.4)/ CFB (0.9)/ HfO$_{2}$ (3.0) \\
FM Stack \verb|#| 3 & sub// Ta (5.0)/ Pt (1.0)/ Ir (0.4)/ CFB (1.0)/ HfO$_{2}$ (3.0) \\
 \hline
\end{tabular}
\caption{\label{tab_fm_material_film}FM material stacks used in this study.}
\label{tab_fm_material_film}
\end{table}

\begin{figure}[h!]
\centering\includegraphics[width=16cm]{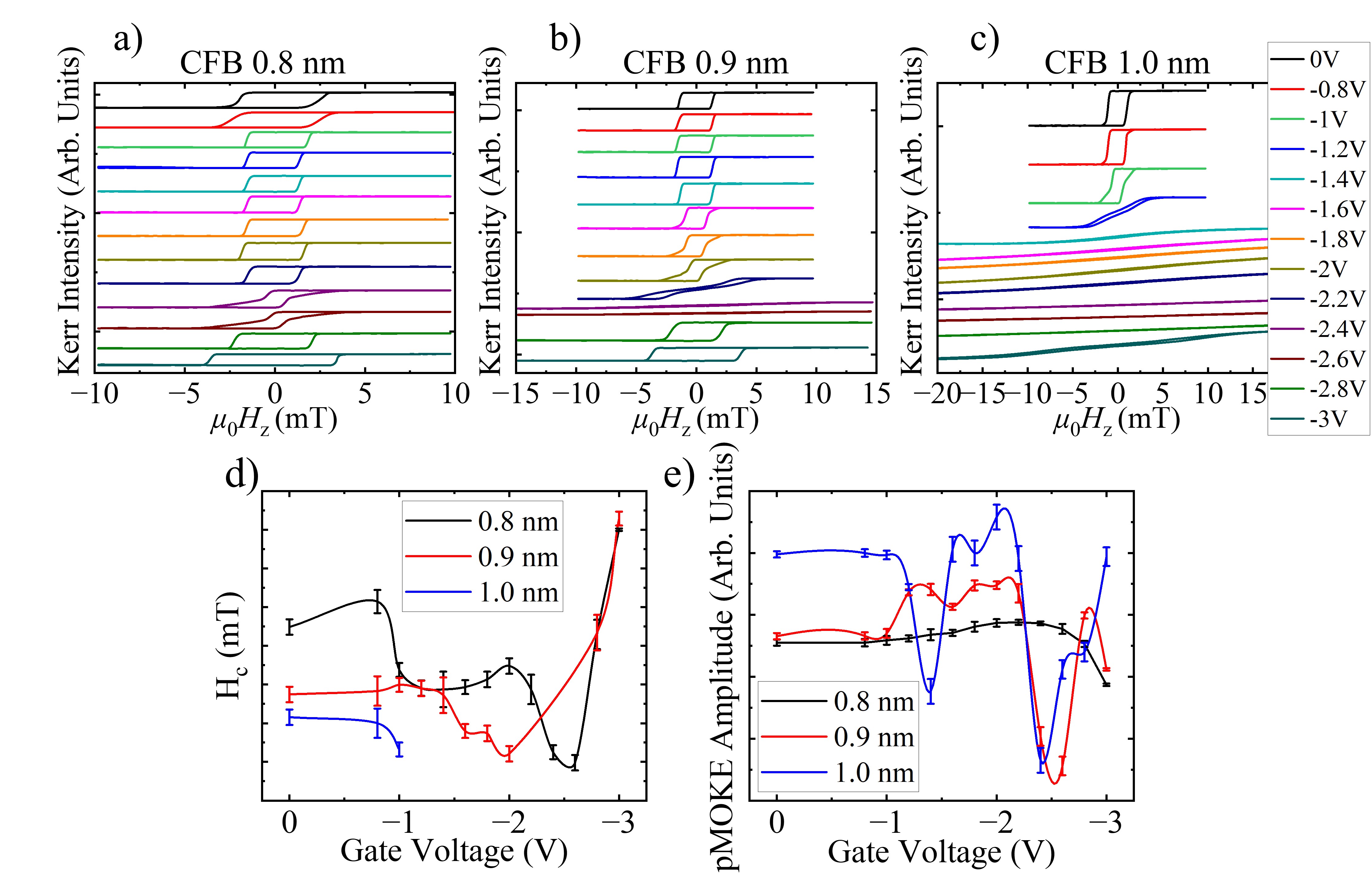}
\caption{\label{fig_Fig3} The non-monotonic behavior of single FM layers as a function of magnetic field was investigated for $t_\text{CFB}$ equal to a) 0.8 nm, b) 0.9 nm, c) 1.0 nm, using pMOKE magnetometry. Hysteresis loops were recorded after each 30-second application of V$_\text{G}$. d) The coercive field and e) the total pMOKE amplitude as a function of $V_\text{G}$ for the single FM stacks.}
\end{figure}

The as-grown state of the three FM stacks exhibits PMA, as indicated by the square hysteresis loops shown in Fig.~\ref{fig_Fig3}a, b and c, corresponding to $0$ V (black curve).
For the 0.8 nm CFB stack, $H_\text{int}$ gradually decreases with the application of $V_\text{G}$, leading to a reduction in the PMA.
This behavior is consistent with the moderate oxidation of the top FM layer observed in Co/HfO$_2$\cite{Liza} and CoFeB/HfO$_2$\cite{Rohit} systems. At $V_\text{G}$ values of -2.4 V and -2.6 V, the square hysteresis transforms into a more distorted hourglass-like shape, approaching the spin re-orientation transition (SRT). However, at -2.8 V, the hysteresis loop regains the square shape, indicating an enhancement of the PMA in the stack.
For the 0.9 nm CFB stack, $H_\text{int}$ remains unchanged in the low V$_\text{G}$ range until -1.6 V, where the square hysteresis transforms into a distorted sigmoidal curve, approaching the SRT, as observed in the 0.8 nm system.
At $V_\text{G}$ equal to -2.4 V, the stack loses PMA, which is then recovered at -2.8 V.
In the case of the 1.0 nm CFB stack, the square hysteresis transforms into the distorted curve at much lower V$_\text{G}$, losing the PMA already at -1.4 V and showing a moderate recovery of the PMA at -3 V. The plots in Fig.~\ref{fig_Fig3}c and d illustrate the variation of $H_\text{c}$ and the total amplitude of the pMOKE hysteresis loops, respectively, as a function of $V_\text{G}$ for all three stacks. Both $H_\text{c}$ and the amplitude of the hysteresis loops exhibit non-monotonic behavior, reminiscent of the oscillatory response observed in the SyAFM stack when influenced by $V_\text{G}$, shown in Fig.~\ref{fig_syafm}b.

The magneto-ionic response is consistently observed in all three thicknesses of the single FM layers, but it becomes more pronounced in the stack with a CFB of 0.9 nm, which serves as a distinctive feature of a two-phased composite FM\cite{6971565}. The insertion of a 0.4 nm Ir layer beneath the FM layer forms an alloyed phase, significantly affecting the overall switching behavior. The intricate relationship between the switching behavior and magneto-ionic gating can be attributed to the interplay between the anisotropies of the two phases and their corresponding dependencies on the gating process. 
In more detail, in the as-grown state, the composite FM consisting of two phases in the composite FM exhibits simultaneous switching behavior. This behavior is observed as a single switching step in samples with thicknesses of 0.8 nm and 0.9 nm, while in the sample with a thickness of 1 nm, the switching occurs at fields that are in very close proximity. As the negative V$_\text{G}$ is gradually increased, the top phase of the FM undergoes moderate oxidation, leading to a decoupling of the two phases. This decoupling allows the phases to switch at more distinct coercive fields, resulting in the appearance of the distorted sigmoidal curves. Eventually, the system undergoes a loss of PMA, indicated by the hard axis loops, leading to an increase in the in-plane component of the magnetic anisotropy. However, with further oxidation, the out-of-plane component becomes stronger again in the final state. This transition from PMA to in-plane anisotropy and again back to PMA can be linked to the non-monotonic and oscillatory behaviour observed in a similar range of V$_\text{G}$ for the SyAFM stacks.

\subsection{X-ray absorption and photoemission spectroscopy analysis}

To comprehensively investigate the changes in magnetic properties of the SyAFM stack in relation to structural and chemical modifications during the ILG process, we conducted Near-edge X-ray absorption fine structure (NEXAFS) spectroscopy and X-ray photoemission spectroscopy (XPS) measurements. These measurements were performed the RGL-PES station of the Russian-German beamline (RGBL) at the synchrotron-radiation facility BESSY II operated by Helmholtz-Zentrum Berlin.For NEXAFS and XPS, the measurements were carried out at an incident photon angle of 45° and an energy resolution better than E/$\Delta$E = 2000. The NEXAFS spectra were acquired in the total electron yield mode. To calibrate the absolute photon energy scale a gold single crystal's Au 4f7/2 photoelectron line (84.0 eV) was measured.

The NEXAFS spectra recorded at the O \textit{K}-edge, as well as at the Fe and Co \textit{L}$_{2,3}$-edges, of both the as-grown state and the gated state obtained after applying Vg = -3V for 30s, are presented in Fig.~\ref{fig_Fig4}a, b and c. The overall shape of the O \textit{K}-edge spectrum  is consistent with that of HfO$_2$, as reported in previous studies \cite{Ok}. The presence of a distinctive pre-edge feature at approximately 531 eV in the gated state spectrum indicates the existence of Fe - O or Co - O bonds, signifying oxidation of Fe or Co atoms \cite{SINGH201748}. 

The oxidation process is further corroborated by the prominent features observed in the Fe and Co \textit{L}$_{2,3}$-edges spectra, which are characteristic of $\alpha$-Fe$2$O$3$ and CoO, respectively, indicating the presence of Fe$^{3+}$ and Co$^{2+}$ ions \cite{PhysRevB.79.033402,LIU2008435,Linjmr}. The presence of a rich multiplet structure at the Co \textit{L}$_{3}$-edge line in the gated state spectrum reveals the admixture of higher valence states of Co to its metallic state \cite{doi:10.1021/acs.jpcc.7b04325,PhysRevB.64.214422}. The appearance of a secondary peak at higher energies relative to the main metallic component in the Fe \textit{L}$_{2,3}$-edge spectrum of the gated state signifies the co-existence of $\alpha$-Fe$_2$O$_3$ and nearly metallic Fe\cite{Vinogradov,PhysRevB.64.214422}. 
The enhanced intensity of the \textit{L}$_{2,3}$-edge spectra for Fe and Co in the gated state suggests a decrease in the average occupancy of $3d$ states\cite{PhysRevLett.75.152},supporting the existence of Co and Fe atoms with higher valence states.

\begin{figure}[h!]
\centering\includegraphics[width=1\linewidth]{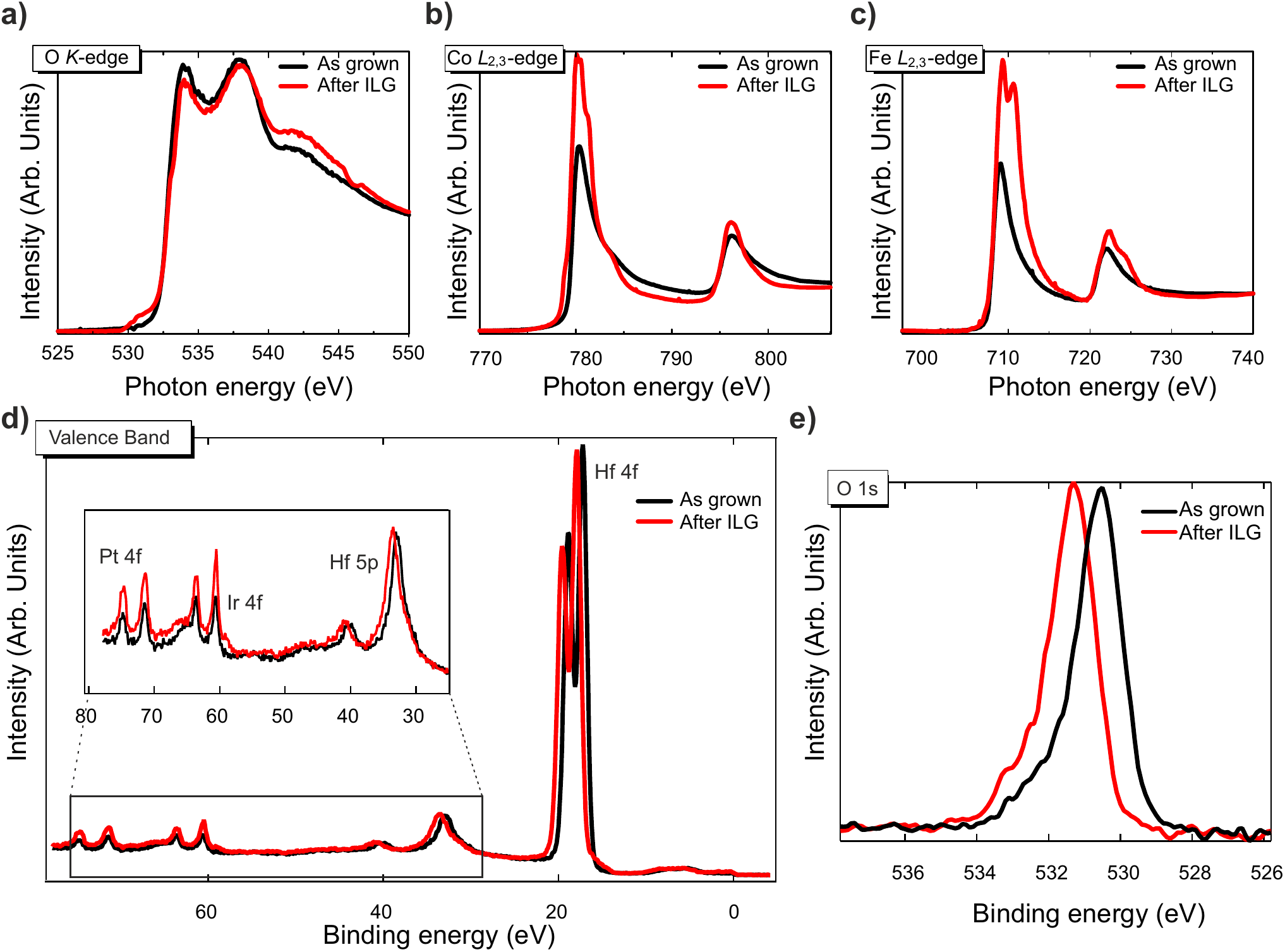}
\caption{\label{fig_Fig4}  Normalized NEXAFS spectra for a) O \textit{K}-edge, b) Fe \textit{L}$_{2,3}$-edges and c) Co \textit{L}$_{2,3}$-edges. XPS spectra of d) Pt, Ir and Hf 4f levels and Hf 5p level, and e) O 1s core level measured at 1000 eV photon energy.}
\end{figure}

The XPS spectra of the Pt 4f and Ir 4f levels remain largely unchanged, as observed in Fig.~\ref{fig_Fig4}d, indicating that oxygen primarily interacts with the top FM layer and may not permeate significantly into the second FM layer. The increase in the intensity could be attributed to the alteration of the chemical environment at the interface\cite{PhysRevLett.75.152}, resulting from the insertion of mobile oxygen species into the metallic stack, causing moderate strain.
The observed displacement of the O and Hf peaks can be attributed to distinct charging dynamics occuring within the oxide layer after the gating process. Through a comprehensive analysis of the XPS and NEXAFS spectra, it becomes evident that mobile oxygen species primarily interact with the top FM layer, with indications suggesting their presence at the Ir/top FM interface within this specific system.

\section{Conclusions}

Our findings reveal an intriguing non-monotonic behavior of the IEC in the SyAFM stacks, providing valuable insight to the interplay between $H_\text{int}$ and $V_\text{G}$.
Through a systematic study, we have identified the key role of the FM layer thickness in shaping the impact of the magneto-ionic effect on the IEC field. The results obtained from the SyAFM systems reveals a non-monotonic, two-oscillation modulation of $H_\text{int}$ as the FM layer thickness increases. Furthermore, our quantitative analysis, based on the single FM layer, suggests the possible formation of a composite FM layer with distinct behaviors. The interplay between the in-plane and out-of-plane anisotropy during the application of $V_\text{G}$ confirms the coexistence of two distinct phases within the single FM layer. These phases exert a significant influence on the switching behavior of the FM system and are likely responsible for the observed magneto-ionic response in the SyAFM systems. The observed interplay is primarily governed by changes in the oxidation states of the FM layer and the chemical environment of the interfaces, as verified by the conducted spectroscopy measurements. These findings are crucial for understanding the role of ILs in controlling the IEC and hold immense potential for advancing the development of highly efficient spintronic devices. The ability to achieve magneto-ionic control in SyAFM structures opens up new possibilities for innovative spintronic device applications.




\begin{acknowledgments}

This project has received funding from the European Union’s Horizon 2020 research and innovation programme under the Marie Skłodowska-Curie grant agreement No 860060 "Magnetism and the effect of Electric Field” (MagnEFi).
This work was partly supported by JSPS Kakenhi (No. 23K13655), as well as funded by the Deutsche Forschungsgemeinschaft (DFG, German Research Foundation)
- TRR 173/2 - 268565370 Spin+X (Projects B02 and A01). Helmholtz-Zentrum Berlin fur Materialien und Energie is acknowledged for provision of access to synchrotron radiation facilities and allocation of synchrotron radiation. S. Kasatikov acknowledges the financial support from the Helmholtz Association (Helmholtz Initiative for refugees) and Helmholtz-Zentrum Berlin.
\end{acknowledgments}

\section*{Author Declarations}
\subsection*{Conflict of interest }
The authors declare no conflict of interest.

\section*{Data Sharing Policy }
The data that support the findings of this study are available from the corresponding author upon reasonable request.





\nocite{*}
\bibliography{bibliography}

\newpage



\section*{Supplementary Material (SM)}
\renewcommand{\thepage}{SM\arabic{page}}
\renewcommand{\thesubsection}{SM\arabic{subsection}}  
\setcounter{page}{1}

\subsection{Effect of magneto-ionic interactions on the magnetic properties of a SyAFM stack}

To reveal the influence of the magneto-ionic effect on the modulation of the IEC strength, we perform ex-situ SQUID measurements, in order to disentangle the impact on different properties, such as saturation magnetization ($M_\text{s}$), uncompensated magnetization ($M_\text{c}$) and IEC field ($H_\text{int}$) as a function of $V_\text{G}$.
The intrinsic IEC parameter is described in Eq.~\ref{eq_Hrkky}, indicating that not only the modulation of interlayer exchange interaction $J_\text{int}$ but also that the modulation of magnetization contributes to the variation of the $H_\text{int}$. Although the previous results demonstrated the AFM-FM transition using the ILG, which implies the change in the $J_\text{int}$ \cite{Yang2018, Guan2021, Kossak}, the magnetization contribution potentially hinders exploiting the genuine behavior of the $J_\text{int}$ under gating only from the $H_\text{int}$ which has been frequently used as the indicator of the modulation of the interlayer exchange interaction\cite{Yang2018_2, Kossak, Ameziane}. Thus, we first determine the $M_\text{c}$, $M_\text{s}$ and $H_\text{int}$, in order to calculate the $J_\text{int}$.

\begin{figure}[h!]
\centering\includegraphics[width=16cm]{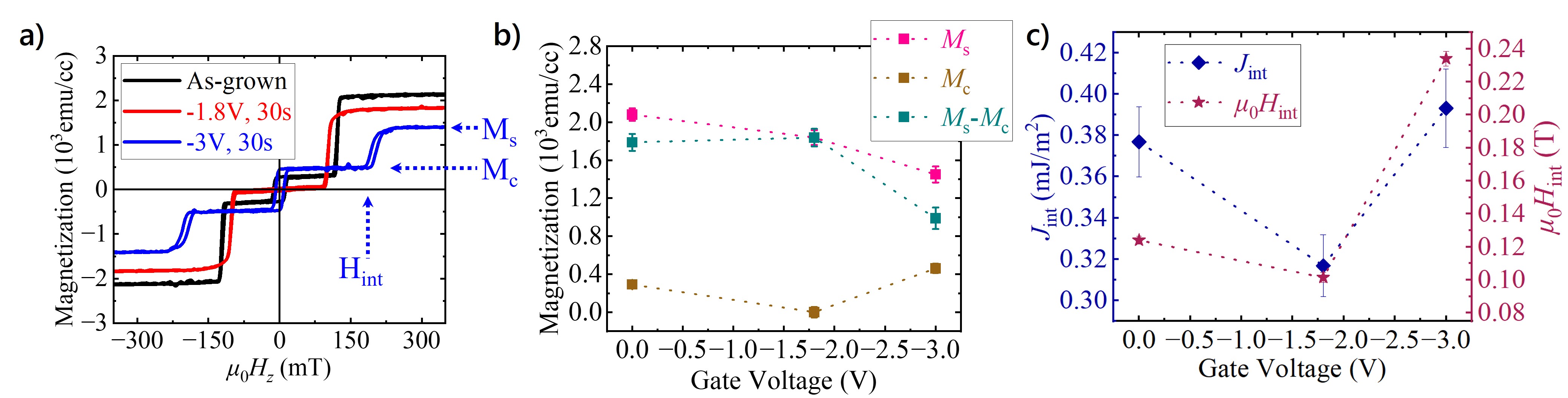}
\renewcommand{\figurename}{FIG. SM}
\setcounter{figure}{0}
\caption{\label{fig_suppl2} Ex-situ SQUID measurements for several $V_\text{G}$ applied on SyAFM stack \#2 with  $t_\text{CFB}$ 0.9 nm as top FM. a) m-H curves, b) saturation magnetization and compensated moment as function of gate voltage, c) IEC strength and coupling field  as a fuction of $V_\text{G}$.}
\end{figure}

\begin{equation} \label{eq_Hrkky}
\mu_0 H_\text{int} = \frac{J_\text{int}}{t_\text{FM} (M_\text{s}-M_\text{c}) }.
\end{equation}

With this in mind, we proceed with the analysis of the findings pertaining to the SyAFM stack \verb|#|2 with 0.9 nm CFB as the top FM layer. In Fig. SM~\ref{fig_suppl2}a, the ${m}$-${H}$ curve shows the spin flip-like AFM-FM transition at $H_\text{int}$, indicating that the two FM layers are antiferromagnetically coupled via the non-magnetic interlayer \cite{Duine2018} at low magnetic fields. By applying negative $V_\text{G}$, corresponding to the migration of mobile oxygen species towards the metallic stack, we observe a decrease in the $M_\text{s}$, as shown in Fig. SM~\ref{fig_suppl2}a, which agrees with the moderate oxidation of the top FM layer, as observed in Co/HfO$_2$ \cite{Liza} and CoFeB/HfO$_2$\cite{Rohit}. The stack reaches 100$\%$ compensation with $V_\text{G}= \minus 1.8$ V for 30 s and a significant increase in the $J_\text{int}$ with $V_\text{G}= \minus 3$ V for 30 s, as shown in Fig. SM~\ref{fig_suppl2}b,c. 
  

Finally, in Fig. SM~\ref{fig_suppl2}c, we observe the non-monotonic modulation of $J_\text{int}$ as a function of $V_\text{G}$ reflecting the modulation of $H_\text{int}$, indicating that the modulation of $M_\text{s}$ can influence the IEC energy quantitatively, but not qualitatively. This intriguing non-monotonic behavior implies not only the change of amplitude but also the phase shift dependence of the oscillation of IEC\cite{PhysRevB.100.014403}, indicating the modulation of the electronic structure by the interplay of structural changes and the strain induced to the stack by the mobile oxygen species or the ions themselves.




\end{document}